\documentclass[aps,prl,twocolumn,groupedaddress]{revtex4}

\usepackage{graphicx, bm, amsfonts, amssymb, amsmath}

\begin{document}

\title{Optimized free energies from bidirectional single-molecule force spectroscopy}

\author{David D. L. Minh}
\email[Electronic Address: ]{daveminh@gmail.com}
\author{Artur B. Adib}

\affiliation{Laboratory of Chemical Physics, NIDDK, National Institutes of Health, Bethesda, Maryland 20892}

\date{\today}

\begin{abstract}
An optimized method for estimating path-ensemble averages using data from processes driven in opposite directions is presented. Based on this estimator, bidirectional expressions for reconstructing free energies and potentials of mean force from single-molecule force spectroscopy---valid for biasing potentials of arbitrary stiffness---are developed.  Numerical simulations on a model potential indicate that these methods perform better than unidirectional strategies.
\end{abstract}

\maketitle
Crooks' path-ensemble average theorem (Eq.~\ref{eq:CPEAT}) encompasses a set of exact results in nonequilibrium statistical mechanics pertinent to systems driven from thermal equilibrium by a time-dependent external potential \cite{Crooks00}. These include Jarzynski's equality \cite{Jarzynski97a} and the Crooks fluctuation theorem \cite{Crooks99}, which relate equilibrium free energy differences to the nonequilibrium work distribution, as well as reweighting relations that allow one to recover arbitrary equilibrium ensemble averages from measurements of driven nonequilibrium processes \cite{Crooks00}. Because of the intimate connection between such processes and molecular force spectroscopy, these theorems have been widely invoked to extract free energies and potentials of mean force (PMFs) from single-molecule pulling experiments \cite{Hummer01a, Hummer05, Liphardt02, Park03, Collin05}.

While formally correct, the practical utility of these relations is limited by the presence of exponential averages of the work, which are dominated by rare events and therefore have notoriously slow convergence properties \cite{Jarzynski97b}. In order to improve their convergence, strategies such as work-weighted trajectory sampling \cite{Sun2003, Ytreberg2004, Oberhofer2005, Oberhofer2008} 
have been proposed.  
Here we suggest another method to accelerate the convergence of these averages: including trajectories from the reverse process in the forward path-ensemble.  This is motivated in part by the observation that the exponential average of the work in the forward process is dominated by those rare trajectories that resemble time-reversed counterparts (``conjugate twins'') of typical trajectories generated by the reverse protocol \cite{Jarzynski06}. Thus, our goals are to construct optimized forward path-ensemble average estimators that explicitly include such trajectories, and apply them to the problem of estimating free energies and potentials of mean force from single-molecule pulling experiments.


The starting point of our analysis is Crooks' path-ensemble average theorem, which relates the forward average of an arbitrary functional $\mathcal{F}=\mathcal{F}[\Gamma]$ of the phase space trajectory $\Gamma = \left\{ q(t), p(t) \right\}$ to its work-weighted average in the reverse process, namely \cite{Crooks00}
\begin{equation}
\left< \mathcal F \right>_F = \left< \hat{\mathcal F} e^{-\beta (W+\Delta F)} \right>_R.
\label{eq:CPEAT}
\end{equation}
In the above, the forward average $\left<...\right>_F$ is an average over all trajectories (path-ensemble average) generated in the forward process, wherein an external parameter (e.g. the position of a harmonic trap in a single-molecule pulling experiment) is driven from the value $A$ to $B$ in $\tau$ units of time after equilibration at $A$, while $\left<...\right>_R$ is a similarly defined average in the reverse direction, from $B$ to $A$. The total work $W[\Gamma]$ accumulated up to the final time $\tau$ is defined in terms of the time-dependent Hamiltonian $H=H(q(t),p(t); t)$ as $W=\int_0^\tau ( \partial H/\partial t ) \, dt$, while $\Delta F = F_B - F_A$ is the free energy difference between the equilibrium states corresponding to the endpoints $A$ and $B$. Finally, the notation $\hat{\mathcal{F}} \equiv \mathcal{F}[\hat{\Gamma}]$ is a shorthand for the value of the functional when evaluated over the time-reversal of $\Gamma$, viz. $\hat{\Gamma} = \left\{q(\tau - t), -p(\tau - t)\right\}$.

By choosing $\mathcal{F}[\Gamma'] = \delta[\Gamma - \Gamma']$ in Eq.~(\ref{eq:CPEAT}) and using the property $W[\hat{\Gamma}] = - W[\Gamma]$, one obtains an identity between the distribution of trajectories in the two directions \cite{Crooks00,crooks98},
\begin{equation}
\rho_F(\Gamma) = e^{ \beta (W - \Delta F)} \, \rho_R(\hat{\Gamma}),
\label{eq:path_CFT}
\end{equation}
where $\rho_F(\Gamma)$ and $\rho_R(\Gamma)$ are the probabilities of observing a particular trajectory $\Gamma$ in the forward and reverse processes, respectively. This result offers a means of achieving the aforementioned goal --- trajectories from the reverse process can indeed be included in the forward path-ensemble when their density is reweighted by $e^{ \beta (W - \Delta F)}$. Our next goal is to  optimally combine direct estimates of $\rho_F(\Gamma)$ from forward processes with indirect estimates obtained from $\rho_R(\Gamma)$ via Eq.~(\ref{eq:path_CFT}); this will be done with the weighted histogram analysis method (WHAM) \cite{Ferrenberg89,Kumar92}.

The objective of WHAM is to find an optimal (i.e. least variance) estimator for a desired probability distribution from a series of independent estimates of biased distributions, where ``biased'' here means that the distribution of interest is related to the remaining ones by a simple reweighting factor. To be specific, given a series of normalized distributions $\rho_i^b(x)$ of a random variable $x$, with $i=1,\ldots,M$, and $M$ unbiasing relations of the form
\begin{equation} \label{eq:WHAM1}
  \rho(x) = f_i(x) \rho_i^b(x),
\end{equation}
where $\rho(x)$ is the distribution of interest and $f_i(x)$ is the unbiasing factor for the $i$-th distribution, the WHAM strategy seeks a linear combination of $M$ independent estimates of $\rho(x)$ obtained from the measured biased distributions $\rho_i^b(x)$ via Eq.~(\ref{eq:WHAM1}), such that its variance $\sigma^2[\rho(x)]$ is minimized.  This results in \cite{Ferrenberg89, Kumar92}
\begin{equation} \label{eq:WHAM2}
  \rho(x) = \frac{\sum_{i=1}^M n_i \, \rho_i^b(x)}{\sum_{i=1}^M n_i \, f_i^{-1}(x)},
\end{equation}
where $n_i$ is the number of samples in the estimate of the $i$-th distribution. (For notational simplicity, here we do not distinguish the exact distribution and its sample estimate). Applied to the problem of estimating $\rho_F(\Gamma)$ from $n_F$ forward and $n_R$ reverse trajectories, Eqs.~(\ref{eq:path_CFT})-(\ref{eq:WHAM2}) give an optimized estimator for the forward probability distribution of trajectories in terms of the measured forward and reverse densities:
\begin{equation} \label{eq:density_optimal}
  \rho_F(\Gamma) = \frac{n_F \rho_F(\Gamma) + n_R \rho_R(\hat{\Gamma})}{n_F + n_R e^{-\beta(W-\Delta F)}}.
\end{equation}

\begin{widetext}
We are now ready to derive the main results of our paper. Taking the average of $\mathcal{F}[\Gamma]$ using the optimized density from Eq.~(\ref{eq:density_optimal}), we obtain the following estimator for the forward path-ensemble average of $\mathcal{F}$:
\begin{equation} \label{eq:F_general}
  \left< \mathcal{F} \right>_F = \left< \frac{n_F \, \mathcal{F}}{n_F + n_R e^{-\beta(W-\Delta F)}} \right>_F +
                                 \left< \frac{n_R \, \hat{\mathcal{F}}}{n_F + n_R e^{\beta(W+\Delta F)}} \right>_R,
\end{equation}
where in the last average we have again used the property that the total work is odd under time-reversal. (An analogous expression for the reverse path-ensemble can be obtained by switching the definitions of forward and reverse.)  This general result forms the basis of our bidirectional method, and different applications can be obtained with suitable choices of $\mathcal{F}$. 

Our first example is concerned with free energy differences, where we choose $\mathcal{F} = e^{-\beta W_0^t}$, with $W_a^b = W_a^b[\Gamma]$ defined as the partial work between times $a$ and $b$ along the trajectory $\Gamma$, i.e. $W_a^b \equiv \int_a^b ( \partial H/\partial t ) \, dt$. (Note that, according to this notation, the total work $W$ coincides with $W_0^\tau$). Invoking Jarzynski's equality $e^{-\beta (F_t - F_A)} = \langle e^{-\beta W_0^t} \rangle_F$ for the l.h.s. of Eq.~(\ref{eq:F_general}), this choice of $\mathcal{F}$ gives
\begin{equation} \label{eq:optimal_Ft}
  e^{-\beta \Delta F_t} = \left< \frac{ n_F \, e^{-\beta W_0^t}}{n_F + n_R e^{-\beta(W-\Delta F)}} \right>_F +
                          \left< \frac{ n_R \, e^{\beta W_{\tau - t}^\tau}} {n_F + n_R e^{\beta(W+\Delta F)}} \right>_R,
\end{equation}
where $\Delta F_t = F_t - F_A$ is the free energy difference between the equilibrium states defined by the Hamiltonians $H(q,p;t)$ and $H(q,p;0)$, and in the last average we have used the property $W_0^t[\hat{\Gamma}] = - W_{\tau - t}^\tau[\Gamma]$. For the particular cases where $t=0$ or $t=\tau$, this result can be rearranged to yield the Bennett Acceptance Ratio (BAR) formula for $\Delta F$ \cite{Bennett76}, as generalized to nonequilibrium processes by Crooks \cite{Crooks00} (for a multistate extension, see \cite{Maragakis2006}). The above equation further generalizes BAR to estimate intermediate free energy differences $\Delta F_t$.  Operationally, when estimating an intermediate free energy difference, we must first estimate $\Delta F$ to use in the r.h.s. of Eq.~(\ref{eq:optimal_Ft}).  This can be accomplished with BAR, which has been shown to be a maximum likelihood estimator of $\Delta F$ \cite{Shirts03}.
\end{widetext}

Free energy differences can also be estimated using a cumulant expansion of Jarzynski's equality \cite{Park03}.  In order to analyze bidirectional data with this approach, one should apply Eq.~(\ref{eq:F_general}) to estimate moments of the work distribution, choosing $\mathcal F = W^n$. This is more rigorous than a method which applies the Crooks fluctuation theorem between states which are not in equilibrium \cite{Kosztin06}.  A bidirectional estimator for the energetic contribution to $\Delta F_t$ can be obtained by choosing $\mathcal F = H(q,p;t) e^{-\beta (W_0^t-\Delta F_t)}$ in Eq.~(\ref{eq:F_general}).  This results in the average energy at time $t$, as was shown in the unidirectional case \cite{Nummela2008}.

In the context of single-molecule pulling experiments, the system is typically driven out of equilibrium by a time-dependent potential $V_t = V(z_t; t)$ acting on a collective coordinate $z_t=z(q(t))$ (e.g. the end-to-end distance of a protein) such that the total Hamiltonian is $H = H_0 + V_t$, where $H_0$ is the (time-independent) Hamiltonian in the absence of the external perturbation. In this case, the free energy difference $\Delta F_t$ involves the equilibrium states of the system corresponding to the potential at $V_t$ and $V_0$. However, one is often more interested in the potential of mean force $G_0(z)$ of the unperturbed Hamiltonian, i.e. in the effective potential dictating the equilibrium distribution of $z$-values in the absence of the external potential.  Although in the limit of sufficiently stiff potentials the free energy difference approaches the PMF \cite{Park03}, this approximation fails for soft springs \cite{Minh2008} such as those used in optical tweezer experiments \cite{Collin05}, in which case one should use more rigorous methods. One approach starts from the observation that the equilibrium distribution of $z$-values in the absence of the external potential (i.e. the unbiased distribution) is given by $\rho_0(z) = C^{-1} e^{\beta V(z;t)} \langle \delta(z - z_t) e^{-\beta W_0^t} \rangle_F$ \cite{Hummer01a, Hummer05}, where $C = \langle e^{-\beta(W_0^t - V_t)} \rangle_F$ is an overall normalization constant, which can be shown to be independent of $t$. With this result in mind, a bidirectional estimator for $\rho_0(z)$ can be obtained from Eq.~(\ref{eq:F_general}) by choosing $\mathcal{F} = \delta(z-z_t) e^{-\beta W_0^t}$. Moreover, since this expression for $\rho_0(z)$ is correct for all times $t$, different estimates of $\rho_0(z)$ can be obtained from different time-slices during the pulling process, and these can in turn be combined according to the WHAM prescription (Eqs.~(\ref{eq:WHAM1}) and (\ref{eq:WHAM2})). Indeed, rewriting the above result for $\rho_0(z)$ in the form of Eq.~(\ref{eq:WHAM1}), viz.
\begin{equation}
  \rho_0(z) = C^{-1} e^{\beta [V(z;t) - \Delta F_t]} \, \left[ \frac{ \langle \delta(z - z_t) e^{-\beta W_0^t} \rangle_F }{ e^{-\beta \Delta F_t} } \right],
\end{equation}
where the factor $e^{-\beta \Delta F_t} = \langle e^{-\beta W_0^t} \rangle_F$ is introduced to normalize the distribution in square brackets, one arrives at the Hummer-Szabo estimator for $\rho_0(z)$ \cite{Hummer01a, Hummer05},
\begin{equation} \label{eq:HSz}
  e^{-\beta G_0(z)} = \frac{ \sum_t \langle \delta(z-z_t) e^{-\beta W_0^t} \rangle_F \, e^{\beta \Delta F_t} }{ \sum_t e^{-\beta [V(z;t) - \Delta F_t]} },
\end{equation}
where $G_0(z) \equiv -\beta^{-1} \ln \rho_0(z)$ is defined up to an additive constant. The above PMF formalism has been extended to account for multiple pulling protocols \cite{Minh2006} and multiple dimensions \cite{Minh2007}.

\begin{widetext}
In order to optimally include trajectories from the reverse perturbation in Eq.~(\ref{eq:HSz}), we choose $\mathcal{F} = \delta(z-z_t) e^{-\beta W_0^t}$ in Eq.~(\ref{eq:F_general}) and substitute the ensuing expression for $\langle \delta(z-z_t) e^{-\beta W_0^t} \rangle_F$ in Eq.~(\ref{eq:HSz}). This leads to our bidirectional PMF estimator:
\begin{equation} \label{eq:Gz}
  e^{-\beta G_0(z)} = \frac{\sum_t 
  \left[
     \left< \frac{ n_F \, \delta(z-z_t) e^{-\beta W_0^t}}{n_F + n_R e^{-\beta(W-\Delta F)}} \right>_F +
     \left< \frac{ n_R \, \delta(z-z_{\tau-t}) e^{\beta W_{\tau - t}^\tau}} {n_F + n_R e^{\beta(W+\Delta F)}} \right>_R
  \right]
  e^{\beta \Delta F_t}}{\sum_t e^{-\beta [V(z;t) - \Delta F_t]}},
\end{equation}
where $\Delta F_t$ is estimated via Eq.~(\ref{eq:optimal_Ft}) and $\Delta F = \Delta F_\tau$ via BAR.  (As in WHAM, $\Delta F_t$ can also be estimated self-consistently by iterative cycles of Eq.~(\ref{eq:Gz}) and  numerically integrating $\Delta F_t = \int e^{-\beta\left[G_o(z)+V(z;t)\right]}dz/\int e^{-\beta\left[G_o(z')+V(z';0)\right]}dz'$.)  If we switch the definitions of forward and reverse, then $G_o(z)$ differs by the constant $\Delta F$.
\end{widetext}

To demonstrate these results, we perform Brownian dynamics simulations on a one-dimensional potential whose unperturbed Hamiltonian is $H_0(z) = (5 z^3 - 10 z + 3) z$ (as used by Hummer \cite{Hummer07}).  The time-dependent Hamiltonian is $H(z;t) = H_0(z) + V(z;t)$, with $V(z;t) = k_s (z - \bar{z}(t))^2/2$ and $k_s$ chosen as 15.  In the forward direction, the center of the potential $\bar{z}(t)$ is linearly varied from -1.5 to 1.5 over 750 steps; it is varied from 1.5 to -1.5 in the reverse direction.  Before pulling, trajectories are equilibrated for 100 steps.  Dynamics are run with a diffusion coefficient D = 1, temperature parameter $\beta = 1$, and time step $\Delta t = 0.001$.  Work is calculated with the discrete formula $W_a^b = \sum_{t=a}^{b-\Delta t} [H(z(t+\Delta t);t+\Delta t) - H(z(t+\Delta t);t)]$.

\begin{figure}[h]
\begin{center}
\includegraphics{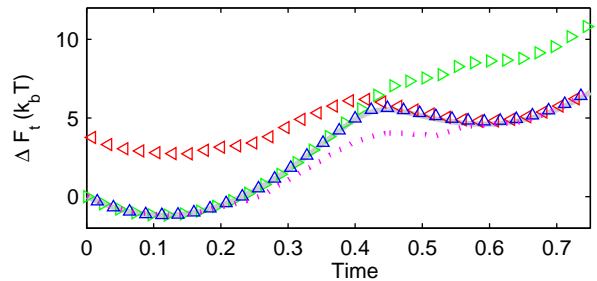}
\caption{\label{fig:Ft}
(Color online) Comparison of $\Delta F_t$ estimators: Jarzynski's equality applied to 500 forward (rightward triangles) or reverse pullings (leftward triangles, time reversed so that $\Delta F_t = \Delta F$ at $t=0.75$); our optimized estimator, Eq.~(\ref{eq:optimal_Ft}) (upward triangles) and Eq.~(16) of Ref.~\cite{Chelli08} (dotted line) applied to 250 pullings in each direction.  The exact $\Delta F_t$, calculated by applying Gauss-Kronrod quadrature in MATLAB 7.5 to numerically integrate $\int e^{-\beta H(z;t)} dz$ between $\bar{z}(t) - 5 < z < \bar{z}(t) + 5$, is shown as a shaded line.
}
\end{center}
\end{figure}

For $\Delta F_t$ estimates on this model system, our bidirectional strategy outperforms existent methods (Fig. \ref{fig:Ft}).  Unidirectional estimates of $\Delta F_t$ based on Jarzynski's equality are markedly biased as the states are further perturbed from the starting equilibrium state.  Chelli and coworkers have also developed an asymptotically correct bidirectional estimator that reduces to BAR at the end states (Eq.~(16) in \cite{Chelli08}). However, their derivation is limited to deterministic systems, and although we have empirical evidence that their estimator approaches the correct $\Delta F_t$ for Brownian simulations in the limit of a large number of trajectories (data not shown), it leads to a more pronounced bias than Eq.~(\ref{eq:optimal_Ft}) for the simulations under the above conditions (Fig.~\ref{fig:Ft}).  We suspect that this slower convergence is due to the use of an unoptimized Jarzynski estimator (see Eq.~(7) in \cite{Chelli08}), but since a general derivation of their results is not yet available, it is presently difficult to verify this as the source of discrepancy between these two estimators, and we leave this question for future investigations.

\begin{figure}[h]
\begin{center}
\includegraphics{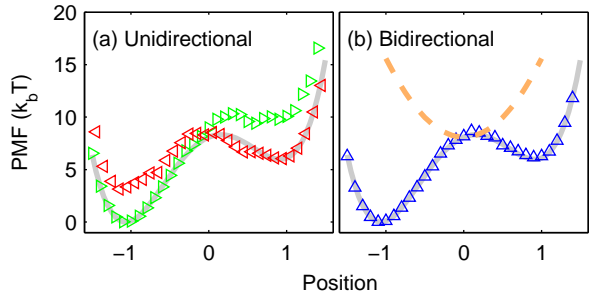}
\caption{\label{fig:PMFs}
(Color online) Comparison of PMF estimators: (a) Hummer and Szabo's method, Eq.~(\ref{eq:HSz}), applied to 500 forward (rightward triangles) or reverse (leftward triangles) pullings.  (b) Our estimator, Eq.~(\ref{eq:Gz}), applied to 250 forward and 250 reverse pullings (upward triangles).  The shaded line is the exact PMF. PMFs from forward and bidirectional data are shifted to align with the exact PMF at $z=-1.25$; for the reverse, they are aligned at $z = 1.25$.  In (b), the harmonic potential used in our pullings is shown as a dashed line.
}
\end{center}
\end{figure}

Our PMF reconstruction methods also compare favorably with unidirectional methods (Fig. \ref{fig:PMFs}).  As with $\Delta F_t$, reconstructed PMFs from separate forward and reverse processes increasingly overestimate the PMF farther from region sampled by the original state.  In contrast, our bidirectional formula, Eq.~(\ref{eq:Gz}), optimally combines the data to reduce this bias.  As the method of Chelli and coworkers for PMF reconstruction requires a stiff-spring assumption, it is not applicable here.

In summary, building on the observation that the convergence of Jarzynski's nonequilibrium work average is dominated by time-reversed counterparts of trajectories generated via the reverse process \cite{Jarzynski06}, we have introduced a formula that optimally includes such trajectories in generic nonequilibrium path-averages (Eq.~(\ref{eq:F_general})). As an application of this result, we have derived a bidirectional estimator for free energy differences in terms of nonequilibrium measurements of work (Eq.~(\ref{eq:optimal_Ft})). Although it reduces to BAR for the special case of endpoint free energy differences $\Delta F$, our formula also allows for the estimation of intermediate values $\Delta F_t$ of the free energy during the switching process. When applied to the problem of estimating potentials of mean force $G_0(z)$ in nonequilibrium force spectroscopy, our methods yield a bidirectional estimator for $G_0(z)$ that optimally combines time-slices from forward and reverse measurements of position and work (Eq.~(\ref{eq:Gz})).  Numerical comparison of our formula with unidirectional estimates based on the Jarzynski equality \cite{Jarzynski97a} or the Hummer-Szabo method \cite{Hummer01a,Hummer05} reveal that our reconstructed free energy differences are of better overall quality than these unidirectional estimators, which are increasingly biased as one drives the system farther away from its original equilibrium state.  It has been noted that faster pullings farther from equilibrium contain less instrument noise and therefore lead to more accurate free energy estimates \cite{Maragakis2007}.  It is thus expected that our bidirectional estimators will further improve the quality of such experimental estimates by appreciably reducing the finite-sample bias most evident in fast pullings.

\section{Acknowledgments}

We thank John Chodera, Gavin Crooks, Gerhard Hummer, Christopher Jarzynski, and Attila Szabo for helpful discussions. This research was supported by the Intramural Research Program of the NIH, NIDDK.


\end{document}